# Entanglement distillation for atomic states via cavity QED

Zhuo-Liang Cao[*], Ming Yang

Department of Physics, Anhui University, Hefei, 230039, PRChina

**Abstract**

In this paper, we proposed a physical scheme to concentrate the non-maximally entangled atomic pure states via cavity QED by using atomic collision in a far-off-resonant cavity. The most distinctive advantage of our scheme is that there is no excitation of cavity mode during the distillation procedure. Therefore the requirement on the quality of cavity is greatly loosened.

---

[*] E-mail address: caoju@mars.ahu.edu.cn

# I    Introduction

Entanglement is a unique feature of quantum mechanics. After Einstein-Podolsky-Rosen's famous work[1], ones begin to seek a perfect understanding of the nature of quantum entanglement. Several years' discussion between prominent scientists in the scientific community has proven the self-consistentance of quantum mechanics. Thereafter, more and more attentions are focused on the non-locality of entanglement. Because of this non-local feature, quantum entanglement has been widely used in quantum information field, such as quantum teleportation[2-3], quantum superdense-coding[4-5], quantum cryptography [6-7], and so on.

To achieve a ideal quantum communication[2,6], maximally entangled states must be distributed among distant users. Because the local operation and classical communication(LOCC) can not generate entanglement between far distant systems[8], the entangled states must be prepared initially in one location and sent to different users. So the distribution of entanglement becomes the vital step in the realization of quantum communication[2,6]. But no one quantum operation is perfect, and no one transmission channel is free of noise. The degradation of entanglement is unavoidable during storing and transmission processes. Therefore, the entangled states shared by different users, are usually non-maximally entangled pure states or the more general case, mixed entangled states. To overcome this problem, several entanglement purification schemes have been proposed[9-17].

For the non-maximally entangled pure states, the process is usually termed as entanglement concentration or distillation[10], and the process dealing with the mixed states usually named as entanglement purification[9]. Entanglement distillation(concentration) is to concentrate a smaller number of maximally entangled states from a large number of non-maximally entangled states by LOCC[10], but the entanglement purification can only increase the entanglement of mixed states by LOCC[9]. For total system, the entanglement hasn't increased, which is in accordance with the result of Ref[8].

Bennett et al have presented the first entanglement concentration scheme[10], where the entanglement of the total system is transferred to the smaller number of entangled pairs. Recently, S. Bose et al proposed another physical scheme to concentrate maximally entangled polarization states of photons using entanglement swapping[17]. Experimental entanglement distillation schemes have also been proposed for the non-maximally entangled polarization states of photons[15,18].

We find that the distillation schemes for entangled polarization states of photons have been researched intensively, and there is a rapid progress in it. But there exist a few distillation schemes for the atomic states. In experiment, atoms are the optimal carrier of quantum information for the quantum computer. So the study on atomic states is of practical significance.

Cavity QED provides a effective tool to deal with the problem on atom and cavity[19-22]. Most of the previous cavity QED schemes use the cavity as the memory of quantum information initially encoded in atomic states. So the decoherence of cavity field becomes one of the main obstacles for the practical implementation of quantum information processing in cavity QED. Recently, Zheng et al have proposed a more effective scheme, which can implement quantum information without the real energy exchange between atom and cavity. In the scheme, the cavity field is only virtually excited during the process, therefore the requirement on the quality of cavity is greatly loosened[23].

GHZ state[24] and W state[25] are two different classes of entangled states for multiparticle system. They all play a important role in quantum communication[26, 27]. So it is necessary and also important to discuss the distillation of them[28, 29].

In this paper, we present a feasible entanglement distillation scheme to concentrate maximally entangled states from non-maximally entangled pure states via cavity QED. Being different from our previous schemes[28, 29], the current one uses the far-off-resonant interaction model[23], therefore the requirement on the effective decoherence time of cavity is greatly prolonged. So it is more feasible in experiment.

The paper is organized as follows. Section II discusses the distillation of GHZ class non-maximally entangled states, the distillation of W class non-maximally entangled states is discussed in section III, and the section IV is the conclusion and discussion.

## II  Physical scheme for the distillation of GHZ class states based on the far-off-resonant interaction between atoms and cavity

The main step of our distillation procedure is the interaction between two atoms and a cavity mode. Suppose that the two atoms are sent through the cavity simultaneously, the Hamiltonian for the total system can be expressed as [23]:

$$\hat{H} = \omega a^+ a + \sum_{j=1,2} \omega_j s_{z,j} + \sum_{j=1,2} \varepsilon_j \left( a s_j^+ + a^+ s_j^- \right), \tag{1}$$

where $s_{z,j}$, $s_j^+$ and $s_j^-$ are atomic operators, and $s_{z,j} = \frac{1}{2}\left(|e\rangle_j\langle e| - |g\rangle_j\langle g|\right)$, $s_j^+ = |e\rangle_j\langle g|$, $s_j^- = |g\rangle_j\langle e|$ with $|e\rangle_j$ and $|g\rangle_j$ being the excited and ground states of $j^{th}$ atom respectively. $a^+$ and $a$ denote the creation and annihilation operators of the cavity mode. $\omega_j$ is the transition frequency of $j^{th}$ atom and $\omega$ is the cavity frequency. $\varepsilon_j$ is the coupling constant between $j^{th}$ atom and cavity mode. Here we suppose that the two atoms are identical, then $\varepsilon_1 = \varepsilon_2 = \varepsilon$ and $\omega_1 = \omega_2 = \omega_0$. If the detuning between atomic frequency and cavity mode frequency $\delta = \omega_0 - \omega$ satisfies $\delta \gg \varepsilon$, there is no energy exchange between the atomic system and the cavity. Then the effective Hamiltonian is rewritten as:

$$\hat{H} = \lambda \left[ \sum_{j=1,2} \left( |e\rangle_j\langle e| aa^+ - |g\rangle_j\langle g| a^+ a \right) + \left( s_1^+ s_2^- + s_1^- s_2^+ \right) \right], \tag{2}$$

where the first two terms describe the stark shifts corresponding to photon number, and the rest two terms are the dipole coupling between the two atoms induced by the cavity mode. $\lambda = \frac{\varepsilon^2}{\delta}$. If the cavity mode is prepared in vacuum state, the effective Hamiltonian takes a new form:

$$\hat{H}_{eff} = \lambda \left[ \sum_{j=1,2} |e\rangle_j \langle e| + \left(s_1^+ s_2^- + s_1^- s_2^+\right) \right]. \tag{3}$$

By solving Schrodinger equation, we can get the evolution of different initial states during the interaction time $t$:

$$|e\rangle_1 |e\rangle_2 \xrightarrow{H_{eff}} e^{-i2\lambda t} |e\rangle_1 |e\rangle_2, \tag{4a}$$

$$|e\rangle_1 |g\rangle_2 \xrightarrow{H_{eff}} e^{-i\lambda t} \left(\cos \lambda t |e\rangle_1 |g\rangle_2 - i \sin \lambda t |g\rangle_1 |e\rangle_2 \right), \tag{4b}$$

$$|g\rangle_1 |e\rangle_2 \xrightarrow{H_{eff}} e^{-i\lambda t} \left(\cos \lambda t |g\rangle_1 |e\rangle_2 - i \sin \lambda t |e\rangle_1 |g\rangle_2 \right), \tag{4c}$$

$$|g\rangle_1 |g\rangle_2 \xrightarrow{H_{eff}} |g\rangle_1 |g\rangle_2. \tag{4d}$$

Now let us consider the detailed distillation procedure.

Firstly, we will discuss the two-atom entangled states case. Suppose that the non-maximally entangled state is in the form:

$$|\Psi_{12}\rangle = a|e_1\rangle|e_2\rangle + b|g_1\rangle|g_2\rangle, \tag{5}$$

where $|a|^2 + |b|^2 = 1, |a| > |b|$. Then atom 1 belongs to Alice, and Bob has access to atom 2.

To get a maximally entangled state from the state in equation (5), an auxiliary atom initially prepared in ground state($|g_a\rangle$) and a cavity initially prepared in vacuum state($|0\rangle$) have to be introduced in the location of Alice or Bob. Without loss of generality, we assume that they are all in Alice's location, and the auxiliary atom and the two entangled atoms are identical.

Alice will let the atom 1 and auxiliary atom through the cavity in the large detuning limit simultaneously. According to the above mentioned model, the system will undergo the following evolution:

$$\begin{aligned}&\left(a|e_1\rangle|e_2\rangle + b|g_1\rangle|g_2\rangle\right)|g_a\rangle \\ &\xrightarrow{U(t)} \left(e^{-i\lambda t} a \cos \lambda t |e_1\rangle|e_2\rangle + b|g_1\rangle|g_2\rangle\right)|g_a\rangle \\ &\quad - i e^{-i\lambda t} a \sin \lambda t |g_1\rangle|e_2\rangle|e_a\rangle\end{aligned} \tag{6}$$

After interaction time $t = \dfrac{1}{\lambda} \arccos \dfrac{|b|}{|a|}$, atoms are flying out of the cavity. Then

Alice will detect the auxiliary atom. If the atom is still in ground state, the two-atom non-maximally entangled state has been concentrated into a maximally entangled state:

$$\sqrt{2}b\left(\frac{1}{\sqrt{2}}(|e\rangle_1|e\rangle_2+|g\rangle_1|g\rangle_2)\right), \tag{7}$$

and the successful probability is $P_{succ}=2|b|^2$. If the atom is detected in excited state, the distillation procedure fails, and the failure probability is $P_{fail}=|a|^2-|b|^2$. The total procedure is depicted in Fig.1.

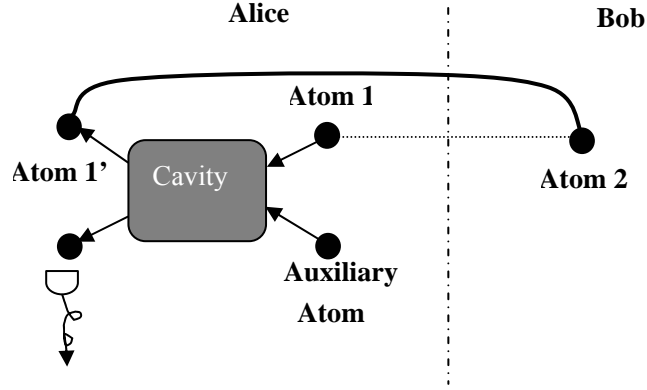

Fig.1. The schematic diagram of the distillation process for 2-atom non-maximally entangled state. The broken line denotes the entanglement of the initial state, and the bold line denotes the entanglement of the concentrated state.

Next, we will consider the 3-atom entangled state case. Suppose the state to be purified is the GHZ class state:

$$|\Psi_{123}\rangle = a|e_1\rangle|e_2\rangle|e_3\rangle + b|g_1\rangle|g_2\rangle|g_3\rangle, \tag{8}$$

where $|a|^2+|b|^2=1, |a|>|b|$, and the three distant users(Alice, Bob and Charlie) have access to the three atoms(1, 2 and 3) respectively. Just like the two-atom case, the auxiliary atom and cavity are introduced at the location of Alice. The auxiliary atom is still prepared in ground state and the cavity is still in vacuum state. At Alice's location, atom 1 and the auxiliary atom will be sent through the cavity simultaneously. According to the same model, the evolution of the total system can be expressed as:

$$\begin{aligned}(a|e_1\rangle|e_2\rangle|e_3\rangle + b|g_1\rangle|g_2\rangle|g_3\rangle)|g_a\rangle \\ \xrightarrow{U(t)} (e^{-i\lambda t}a\cos\lambda t|e_1\rangle|e_2\rangle|e_3\rangle + b|g_1\rangle|g_2\rangle|g_3\rangle)|g_a\rangle \\ - ie^{-i\lambda t}a\sin\lambda t|g_1\rangle|e_2\rangle|e_3\rangle|e_a\rangle\end{aligned} \quad (9)$$

After the two atoms flying out of the cavity, Alice will detect the auxiliary atom. If we select the interaction time $t = \frac{1}{\lambda}\arccos\frac{|b|}{|a|}$, the distillation procedure will succeed with probability $P_{succ} = 2|b|^2$ corresponding to the measurement result $|g_a\rangle$, or fail with probability $P_{fail} = |a|^2 - |b|^2$ corresponding to the measurement result $|e_a\rangle$.

As a straight extension of our scheme, this scheme can also concentrate the multi-atom GHZ class state:

$$|\Psi_{12\cdots 2n}\rangle = a|e_1\rangle|e_2\rangle\cdots|e_{2n}\rangle + b|g_1\rangle|g_2\rangle\cdots|g_{2n}\rangle. \quad (10)$$

The *2n* atoms are distributed among *2n* distant users including Alice. Then the local operations on the entangled atom 1, the auxiliary atom and the cavity at Alice's location can extract maximally entangled states from states(10). The model is all the same to 2-atom and 3-atom cases. After analysis, we get the result: $P_{succ} = 2|b|^2$ and $P_{fail} = |a|^2 - |b|^2$.

We find that we only need to carry local operations at one location, and need no more than one auxiliary atom and a cavity as auxiliary system for the 2-atom, 3-atom and multi-atom cases. So the scheme is a rather simple one. And the successful probability is only dependent on the smallest coefficient of the superposition state to be concentrated. Thus, we have discussed the distillation for the GHZ class state. But W class state is another class state different from the GHZ class state, there is necessity to consider the distillation of W class state.

## III  Physical scheme for the distillation of W class states based on the far-off-resonant interaction between atoms and cavity

We can suppose that the non-maximally entangled W state($|W'\rangle$) is in the form:

$$|W'_3\rangle = a|e_1\rangle|g_2\rangle|g_3\rangle + b|g_1\rangle|e_2\rangle|g_3\rangle + c|g_1\rangle|g_2\rangle|e_3\rangle, \tag{11}$$

where $|a|^2 + |b|^2 + |c|^2 = 1$, $|a| \geq |b| \geq |c|$. These atoms(1, 2 and 3) are distributed among three distant users (Alice, Bob and Charlie) respectively.

Unlike the distillation schemes for the GHZ class states, the distillation scheme for W class states need more than one auxiliary atom and one cavity. For three-atom W class state, we need to introduce an auxiliary atom($a_1$) and a cavity($c_1$) in Alice's location, and another auxiliary atom($a_2$) and cavity($c_2$) in Bob's location. Suppose all the auxiliary atoms and the entangled atoms are identical, and the auxiliary atoms are all prepared in ground state initially.

To concentrate the maximally entangled W state from equation (11), Alice and Bob will carry out same operations, i.e. to send the auxiliary atom and entangled atom into the corresponding cavity initially prepared in vacuum state simultaneously. In the large detuning limit($\delta \gg \varepsilon$), we can adopt the model in the equations (3) and (4). Then the state of the total system will undergo the following evolution:

$$\begin{aligned}
&(a|e_1\rangle|g_2\rangle|g_3\rangle + b|g_1\rangle|e_2\rangle|g_3\rangle + c|g_1\rangle|g_2\rangle|e_3\rangle)|g_{a1}\rangle|g_{a2}\rangle \\
&\xrightarrow{U(t_1)U(t_2)} [e^{-i\lambda t_1}a\cos\lambda t_1|e_1\rangle|g_2\rangle|g_3\rangle + e^{-i\lambda t_2}b\cos\lambda t_2|g_1\rangle|e_2\rangle|g_3\rangle \\
&\quad + c|g_1\rangle|g_2\rangle|e_3\rangle]|g_{a1}\rangle|g_{a2}\rangle \\
&\quad -i(e^{-i\lambda t_1}a\sin\lambda t_1|g_1\rangle|g_2\rangle|g_3\rangle|e_{a1}\rangle|g_{a2}\rangle \\
&\quad + e^{-i\lambda t_2}b\sin\lambda t_2|g_1\rangle|g_2\rangle|g_3\rangle|g_{a1}\rangle|e_{a2}\rangle).
\end{aligned} \tag{12}$$

After interaction times $t_1 = \frac{1}{\lambda}\arccos\frac{|c|}{|a|}$ (Alice) and $t_2 = \frac{1}{\lambda}\arccos\frac{|c|}{|b|}$ (Bob), Alice and Bob will operate a single atom measurement on their auxiliary atoms. After measurement Bob can inform Alice his result via classical communication, so Alice only need one-way classical communication to decide whether the distillation succeeds.. If the joint result is $|g_{a1}\rangle|g_{a2}\rangle$, distillation succeeds with probability $P_{succ} = 3|c|^2$. Otherwise, the distillation procedure fails with probability

$$P_{fail} = 1 - 3|c|^2.$$

In the above discussion, we suppose the auxiliary atoms are initially prepared in ground state. If the auxiliary atoms are initially prepared in excited state, what will happen?

In this case, the auxiliary atom and cavity are introduced by users Bob and Charlie. Then the evolution will take a new form:

$$(a|e_1\rangle|g_2\rangle|g_3\rangle + b|g_1\rangle|e_2\rangle|g_3\rangle + c|g_1\rangle|g_2\rangle|e_3\rangle)|e_{a2}\rangle|e_{a3}\rangle$$

$$\xrightarrow{U(t_2)U(t_3)} (e^{-i\lambda(t_2+t_3)}a\cos\lambda t_2\cos\lambda t_3|e_1\rangle|g_2\rangle|g_3\rangle + e^{-i\lambda(2t_2+t_3)}b\cos\lambda t_3|g_1\rangle|e_2\rangle|g_3\rangle$$
$$+ e^{-i\lambda(t_2+2t_3)}c\cos\lambda t_2|g_1\rangle|g_2\rangle|e_3\rangle)|e_{a2}\rangle|e_{a3}\rangle$$
$$-i(e^{-i\lambda(t_2+t_3)}a\cos\lambda t_2\sin\lambda t_3|e_1\rangle|g_2\rangle|e_3\rangle + e^{-i\lambda(2t_2+t_3)}b\sin\lambda t_3|g_1\rangle|e_2\rangle|e_3\rangle)|e_{a2}\rangle|g_{a3}\rangle$$
$$-i(e^{-i\lambda(t_2+t_3)}a\sin\lambda t_2\cos\lambda t_3|e_1\rangle|e_2\rangle|g_3\rangle + e^{-i\lambda(t_2+2t_3)}c\sin\lambda t_2|g_1\rangle|e_2\rangle|e_3\rangle)|g_{a2}\rangle|e_{a3}\rangle$$
$$-e^{-i\lambda(t_2+t_3)}a\sin\lambda t_2\sin\lambda t_3|e_1\rangle|e_2\rangle|e_3\rangle|g_{a2}\rangle|g_{a3}\rangle. \qquad (13)$$

After interaction times $t_2 = \frac{1}{\lambda}\arccos\frac{|b|}{|a|}$ (Bob) and $t_3 = \frac{1}{\lambda}\arccos\frac{|c|}{|a|}$ (Charlie), Bob and Charlie will measure their own auxiliary atoms respectively. Corresponding to the result $|e_{a2}\rangle|e_{a3}\rangle$, the atoms 1, 2 and 3 are left in maximally entangled W state with successful probability $P_{succ} = \frac{3|b|^2|c|^2}{|a|^2}$. Corresponding to the results $|e_{a2}\rangle|g_{a3}\rangle$ and $|g_{a2}\rangle|e_{a3}\rangle$ the distillation for W state fails, and the three atoms are left in unentangled state for three atoms. But, for the result $|e_{a2}\rangle|g_{a3}\rangle$, atoms 1, 2 are left in two-atom maximally entangled state with probability $P_{succ} = \frac{2(|a|^2-|c|^2)|b|^2}{|a|^2}$, and for the result $|g_{a2}\rangle|e_{a3}\rangle$, atoms 1, 3 are left in 2-atom Bell state with probability $P_{succ} = \frac{2(|a|^2-|b|^2)|c|^2}{|a|^2}$. Corresponding to the result $|g_{a2}\rangle|g_{a3}\rangle$, the distillation procedure fails thoroughly $P_{fail} = \frac{(|a|^2-|b|^2)(|a|^2-|c|^2)}{|a|^2}$.

As a straight extension, we will discuss the distillation of *N*-atom W class state. The *N* atoms are distributed among *N* users. Suppose the *N* atoms are in the state:

$$|W'_N\rangle = c_1|e_1\rangle|g_2\rangle\cdots|g_N\rangle + c_2|g_1\rangle|e_2\rangle|g_3\rangle\cdots|g_N\rangle + \cdots + c_N|g_1\rangle\cdots|g_{N-1}\rangle|e_N\rangle, \quad (14)$$

where $|c_1|+|c_2|+\cdots|c_N|=1$. *N*-1 auxiliary atoms and *N*-1 cavities will be introduced. That is to say there is an auxiliary atom and a cavity in every location except for a special user(defined as $j^{th}$ user below). The cavities are all prepared in vacuum state. Then we will discuss two different cases.

The first(it will be denoted by AUX(e) in the text) is that all the auxiliary atoms are initially prepared in excited state, and $|c_j|\rangle|c_i|$ $(i=1,2,\cdots,N.\ i\neq j)$. Then in the large detuning limit, the evolution of the system is decided by the above mentioned model[30]. After interaction times $t_i = \frac{1}{\lambda}\arccos\frac{|c_i|}{|c_j|}$, there will be a single atom measurement on the auxiliary atom at every location. If the result is $|e_{a_1}\rangle|e_{a_2}\rangle\cdots|e_{a_{j-1}}\rangle|e_{a_{j+1}}\rangle\cdots|e_{a_N}\rangle$, the *N* atoms are left in the maximally entangled W state with probability $P = N\frac{|c_1|^2|c_2|^2\cdots|c_{j-1}|^2|c_{j+1}|^2\cdots|c_N|^2}{|c_j|^{2(N-2)}}$. If the result is $|g_{a_1}\rangle\cdots|g_{a_{i-1}}\rangle|e_{a_i}\rangle|g_{a_{i+1}}\rangle\cdots|g_{a_{j-1}}\rangle|g_{a_{j+1}}\rangle\cdots|g_{a_N}\rangle$, the i$^{th}$ atom and the j$^{th}$ atom collapse into a two-atom maximally entangled state with probability $P = 2|c_j|^2|\sin\lambda t_1|^2\cdots|\sin\lambda t_{i-1}|^2|\cos\lambda t_i|^2|\sin\lambda t_{i+1}|^2\cdots|\sin\lambda t_{j-1}|^2|\sin\lambda t_{j+1}|^2\cdots|\sin\lambda t_N|^2$. If the result is: $|e_{a_1}\rangle\cdots|e_{a_{i-1}}\rangle|g_{a_i}\rangle|e_{a_{i+1}}\rangle\cdots|e_{a_{j-1}}\rangle|e_{a_{j+1}}\rangle\cdots|e_{a_N}\rangle$, the atoms *1, 2, …, i-1, i+1…, N* are left in a *N-1*-atom maximally entangled W state with probability $P = (N-1)|c_j|^2|\cos\lambda t_1|^2\cdots|\cos\lambda t_{i-1}|^2|\sin\lambda t_i|^2|\cos\lambda t_{i+1}|^2\cdots|\cos\lambda t_{j-1}|^2|\cos\lambda t_{j+1}|^2\cdots|\cos\lambda t_N|^2$.

For the second case(it will be denoted by AUX(g) in the text), we suppose that all the auxiliary atoms are in ground state, and $|c_j|\langle|c_i|$ $(i=1,2,\cdots,N.\ i\neq j)$. Then adopting the same operations, we get the evolution of the system[30].

After interaction times $t_i = \frac{1}{\lambda}\arccos\frac{|c_j|}{|c_i|}$, the measurement on auxiliary atoms will be operated by the *N-1* users. we can get the *N*-atom maximally entangled W state with probability $P_{succ} = N|c_j|^2$ corresponding to the result $|g_{a_1}\rangle|g_{a_2}\rangle\cdots|g_{a_{j-1}}\rangle|g_{a_{j+1}}\rangle\cdots|g_{a_N}\rangle$. Otherwise, the distillation fails with probability $P_{fail} = 1 - N|c_j|^2$.

From above analysis, we conclude that, in the first case where the auxiliary atoms are all initially prepared in excited state, the *N*-atom maximally entangled W state can be concentrated from the *N*-atom non-maximally entangled W class state probabilistically. When the distillation procedure fails for the *N*-atom maximally entangled W state, there is a fractional probability to obtain *N*-1-atom maximally entangled W state. But, for the second case where all the auxiliary atoms are in the ground state initially, the scheme can only concentrate the *N*-atom maximally entangled W state.

From the expression of the successful probability of obtaining the *N*-atom entangled W state, we get that $P_{succ(e)} < P_{succ(g)}$, where the $P_{succ(e)}$ denotes the probability of obtaining the *N*-atom entangled W state in the case where the auxiliary atoms are all in the excited state initially, and $P_{succ(g)}$ denotes the successful probability in the case where the auxiliary atoms are all in the ground state initially. It can be understood easily, in the case AUX(g) the entanglement is all concentrated into the *N*-atom maximally entangled W state, but in the second case AUX(e), the entanglement is concentrated into different maximally entangled states, such as *N*-atom maximally entangled W state, *N*-1-atom maximally entangled W state, and two atoms Bell state.

## IV    Conclusion and discussion

In this paper, we proposed an entanglement distillation scheme for non-maximally entangled pure GHZ class state and non-maximally entangled pure W class state. The scheme is based on the far-of-resonant interaction between atoms and a cavity mode, which overcomes the difficulty in obtaining the resonant condition required in other distillation schemes[28, 29]. Here the cavity is prepared in the vacuum state, but the auxiliary atoms can prepared in two different states, ground state or excited state. For the GHZ class state, the two cases give the same result. For the W class state, the two cases give two different results.

In addition, there must be one-way classical communication between different users who have access to auxiliary atom and cavity in the distillation of W class state. But in the GHZ case, the classical communication is not needed. In this sense we can get some reasons why there is a more robust entanglement in W state than in GHZ state.

The most distinctive advantage of our scheme are 1': there is no transfer of quantum information between atoms and cavity, and 2': the cavity is only virtually excited during the distillation procedure. Therefore the requirement on the quality of cavity is greatly loosened. The far-off-resonant condition can be more easily obtained than the resonant one. The disadvantage of the scheme is that the difficulty of sending two atoms into cavity simultaneously. In experiment, it is difficulty to synchronize the two atoms. In addition, after flying out of cavity, there is still a problem about how to distinguish the two atoms.

Recently, the experimental realization of Zheng's proposal has been reported[31]. Our proposal is a feasible scheme, and will be probably realized in the near future.

## Acknowledgements

This work is supported by The Project Supported by Anhui Provincial Natural Science Foundation under Grant No: 03042401 and the Natural Science Foundation of the Education Department of Anhui Province under Grant No: 2002kj026, also by the fund of the Core Teacher of Ministry of National Education under Grant No: 200065.